\documentclass[10pt, conference, compsocconf]{IEEEtran_origin}

\IEEEoverridecommandlockouts

\usepackage{booktabs} % For formal tables
\usepackage[ruled, linesnumbered]{algorithm2e} % For algorithms
\usepackage{xcolor}
\usepackage[normalem]{ulem}
\usepackage{amsmath}
\usepackage{subcaption}
\usepackage{siunitx}
\usepackage{amsthm}
\usepackage{tabularx}
\usepackage{draftcopy}		% Uncomment this line to have the
\usepackage{ifthen}

\newboolean{authnotes}
\setboolean{authnotes}{false}
\ifthenelse{\boolean{authnotes}}
{
\newcommand{\cj}[1]{\footnote{\color{magenta}{\bf Christine: #1}}}
\newcommand{\cl}[1]{\footnote{\color{orange}{\bf Chenguang: #1}}}
\newcommand{\jh}[1]{\footnote{\color{cyan}{\bf Jie: #1}}}
\newcommand{\tk}[1]{\footnote{\color{brown}{\bf Tomasz: #1}}}
\newcommand{\ys}[1]{\footnote{\color{yellow}{\bf Yosef: #1}}}

\usepackage{draftwatermark}
\SetWatermarkLightness{0.95}
\SetWatermarkScale{1}
}
{
\newcommand{\cj}[1]{}
\newcommand{\cl}[1]{}
\newcommand{\jh}[1]{}
\newcommand{\tk}[1]{}
\newcommand{\ys}[1]{}

\usepackage{draftwatermark}
\SetWatermarkText{}
}
\newcommand{\systemname}{\textbf{r}IoT\xspace}
\newcommand{\context}{context\xspace}
\newcommand{\request}{request\xspace}
\newcommand{\action}{action\xspace}

\newcommand{\utility}{utility\xspace}
\newcommand{\device}{device\xspace}
\newcommand{\proposal}{proposal\xspace}

\theoremstyle{definition}
\newtheorem{defn}{Definition}
\definecolor{LightCyan}{rgb}{0.88,1,1}

\let\oldnl\nl% Store \nl in \oldnl
\newcommand{\nonl}{\renewcommand{\nl}{\let\nl\oldnl}}% Remove line number for one line
\begin{document}

\title{\systemname: Enabling Seamless Context-Aware Automation in the Internet of Things}

\author{
    \IEEEauthorblockN{Jie Hua\IEEEauthorrefmark{2}, Chenguang Liu\IEEEauthorrefmark{2}, Tomasz Kalbarczyk\IEEEauthorrefmark{2}, Catherine Wright\IEEEauthorrefmark{3}, Gruia-Catalin Roman\IEEEauthorrefmark{3}, Christine Julien\IEEEauthorrefmark{2}}
    \IEEEauthorblockA{\IEEEauthorrefmark{2}Department of Electrical and Computer Engineering, University of Texas at Austin
    \\\{mich94hj, liuchg, tkalbar, c.julien\}@utexas.edu}
    \IEEEauthorblockA{\IEEEauthorrefmark{3}Department of Computer Science, University of New Mexico
    \\\{wrightc, gcroman\}@unm.edu}
}

% \author{\IEEEauthorblockN{Jie Hua, Chenguang Liu, Tomasz Kalbarczyk, Christine Julien}
% \IEEEauthorblockA{Department of Electrical and Computer Engineering,\\ University of Texas at Austin\\
% \{mich94hj, liuchg, tkalbar, c.julien\}@utexas.edu}
% \and
% \IEEEauthorblockN{Catherine Wright, Gruia-Catalin Roman}
% \IEEEauthorblockA{Department of Computer Science,\\
% University of New Mexico\\
% \{wrightc, gcroman\}@unm.edu}
% }

\maketitle
% \thispagestyle{plain}
% \pagestyle{plain}
%\note{The deadline for MASS 2019: Jun 1, 2019 07:59:59 EDT}

\begin{abstract}
%The difficulties of supporting user-configurable automation are due to the complexity of various dynamic home environments
% {\color{red}Abstract goes here... Note: we can't use the name Warble since the review is double blind}
Advances in mobile computing capabilities and an increasing number of Internet of Things (IoT) devices have enriched the possibilities of the IoT but have also increased the cognitive load required of IoT users. Existing context-aware systems provide various levels of automation in the IoT. Many of these systems adaptively take decisions on how to provide services based on assumptions made {\em a priori}. The approaches are difficult to personalize to an individual's dynamic environment, and thus today's smart IoT spaces often demand complex and specialized interactions with the user in order to provide tailored services. We propose \systemname, a framework for seamless and personalized automation of human-device interaction in the IoT. \systemname leverages existing technologies to operate across heterogeneous devices and networks to provide a one-stop solution for device interaction in the IoT. We show how \systemname exploits similarities between contexts and employs a decision-tree like method to adaptively capture a user's preferences from a small number of interactions with the IoT space. We measure the performance of \systemname on two real-world data sets and a real mobile device in terms of accuracy, learning speed, and latency in comparison to two state-of-the-art machine learning algorithms.

\end{abstract}


\setlength{\textfloatsep}{6pt}% Remove \textfloatsep

% \keywords{}

\section{introduction}\label{sec:introduction}
%\note{This paragraph states the need for IoT automation and the fundamental meaning of context-awareness in automation.}
With recent technology advances made in the Internet-of-Things (IoT), there is a growing number of smart devices helping to build the many smart-* scenarios that people have long envisioned~\cite{weiser1991computer}.
In scenarios like smart-homes and smart-offices, the plethora of these new devices has created many possibilities for automating daily tasks. At the same time, new challenges arise; a particular challenge to note is that applications demand responsive and intelligent approaches to leverage context~\cite{dey2001understanding} in IoT environments. In this work, we address a fundamental piece of this challenge: automating  human-device interaction, by asking a simple yet unsolved question: {\it how can contextual information be leveraged to make IoT device interaction more seamless and personalized}?

% {\em seamless}, {\em context-aware}, and {\em personalized} human-device interaction in the IoT so that the user does not need to manually set up every device or reconfigure the available devices as a result of changes in the environment (e.g., when an occupant moves a floor light to another room).}

%\note{This paragraph make the idea of seamless and personalization concrete.}

To make {\em seamlessness} and {\em personalization} concrete, consider a smart home system, embedded with sensors and actuators. Smart lights adjust lighting based on indoor illumination; a smart coffee maker automatically starts coffee when the user wakes up. While these individual applications enable some automation, they do not achieve the full vision of disappearing computer~\cite{weiser1991computer}. One gap that remains is directly related to abstractions for user interaction chosen by manufacturers~\cite{brush2018smart, clark2017devices}. Simply put, interactions are not {\em seamless}. At setup, a user usually needs to connect the devices based on mac addresses, name each device, and remember them. To interact with devices, the user either scripts the behavior in advance in arbitrary {\em computer-friendly} languages (e.g., IFTTT~\cite{ovadia2014automate}, Hydra~\cite{eisenhauer2010hydra}) or must recall the name defined at setup and issue commands like ``set the light over the stove to bright''. Neither the scripted behavior nor the tailored commands provide a {\em seamless} interaction paradigm. We argue that a truly {\em seamless} IoT world will allow the user to interact with devices using simple and generic instructions like ``turn the light on''.

On the other hand, a key selling point of IoT applications is the {\em personalization} they enable by allowing users to customize the configuration with personal preferences. While such features are common, they are often limited and contrived~\cite{black2014internet}. For instance, although the smart coffee machine may allow the user to configure a customized time to start the coffee machine every day, such ``personalization'' is under the assumption that the interaction related to this device is based on time. If a user wants to start coffee after returning home from jogging, which may or may not happen every day, the user cannot benefit from the ``smartness'' of the coffee machine. Personalization in modern IoT systems should not require a user to express her preference via manufacturer defined assumptions.

Our solution to {\em seamlessness} and {\em personalization} is through context-awareness. Significant work has been done in context-awareness over the past decade~\cite{gil2016internet,perera2014context, sezer2018context}, supporting better collecting, processing, and reasoning about {\em context}. In the IoT, a user's context can include any information that describes the user's situation, from location and time to ambient conditions or the presence of others~\cite{dourish2004we}. We focus on utilizing collected contextual information to predict the device and associated service(s) that a user needs when the user makes a simple generic request (e.g., ``turn on the light''). Unlike existing solutions, we respond immediately to users-supplied negative feedback and re-attempt the action.

We propose, \systemname, a framework enabling \textbf{r}esponsive and \textbf{r}einforcing automation in IoT service personalization. 
\systemname{} enables context-aware automation by providing a {\em seamless} and {\em low effort} approach to personalizing how IoT services are chosen to support a given request from a user.
In contrast to the common IoT application workflow in which users must exert non-negligible efforts on the process of configuring, labeling, and specifying a device before using it, \systemname incorporates a context learning algorithm that automatically adjusts based on user intentions and the environment. This learning is a continuous process that adaptively evolves a learned model as users change their interaction preferences or the environment changes. 
%In comparison to existing semantic-based automation approaches~\cite{gu2004ontology, maarala2017semantic, meng2016rule}, 
\systemname does not rely on {\em a priori} knowledge and learns a user's intentions only from the history of user interactions. 
%Further, we remain attentive to {\em personalization} by learning a user's interaction patterns\cl{sounds repetitive based on the last sentence} from the context, requests, and feedback.

In summary, \systemname leverages rich contextual information to enable increased automation of user-oriented IoT device interaction. Our key research contributions are:
    \begin{itemize}
        \item We propose a context-aware learning framework, \systemname, for user-oriented IoT device selection.
        \item We incorporate user configurable context abstractions to enable personalization at per device level.
        \item We devise a context-aware algorithm that learns a user's interaction pattern with no {\em a priori} knowledge about the device, space, and user.
        \item We quantitatively evaluate \systemname using two real-world data sets. We show that \systemname has high accuracy and can quickly recover from environmental dynamics.
    \end{itemize} 
    
In Section~\ref{sec:related_works}, we present an overview of the related work and key preliminaries of our proposed approach. We then present an overview of \systemname and its position in an IoT deployment. Section~\ref{sec:riot_framework} presents \systemname in detail, including the underpinning learning algorithms. We evaluate \systemname in Section~\ref{sec:evaluation}, comparing it to two alternative learning algorithms in the context of real-world contextual data. Section~\ref{sec:conclusion} concludes.

\section{Motivation and Related Work}\label{sec:related_works}
We use state of the art middleware in the IoT to motivate the gap that \systemname fills. IoT devices fall into two categories: {\em sensors} and {\em actuators}. Users make requests to actuators to take some action (e.g., turn on the lights, adjust the volume, etc.), while sensors passively collect contextual information. A device can take on both roles simultaneously, e.g., a thermostat can both sense temperature and actuate the temperature set point.

{\bf Motivating Application Scenario.}
Alice is a smart home enthusiast who owns several IoT actuators: a smart lock, lights, security cameras, and a stereo system. She is an early adopter who purchases solutions as they become available, so her devices are from five different manufacturers. Alice also has a networked sensor system throughout her home, provided by yet another manufacturer. The stereo system is her favorite because it supports sophisticated collaboration among all of her speakers to provide ideal sound quality. Alice must figure out how to control all of the devices to satisfy her needs but minimize her overhead in interacting with them.

\begin{figure}[t]
\centering
\begin{subfigure}[b]{.48\columnwidth}
\includegraphics[width=\linewidth]{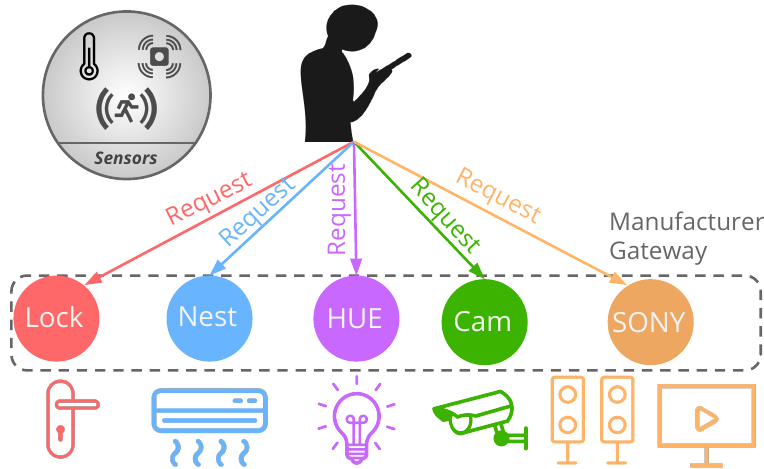}
%https://docs.google.com/drawings/d/1sqB_OM83G0WvekipL6rA3Il31sbXTiQkH34_GcMiVXE/edit
\caption{Manufacturer oriented}
\label{fig:iotManufacture}
\end{subfigure}
\begin{subfigure}[b]{.48\columnwidth}
\includegraphics[width=\linewidth]{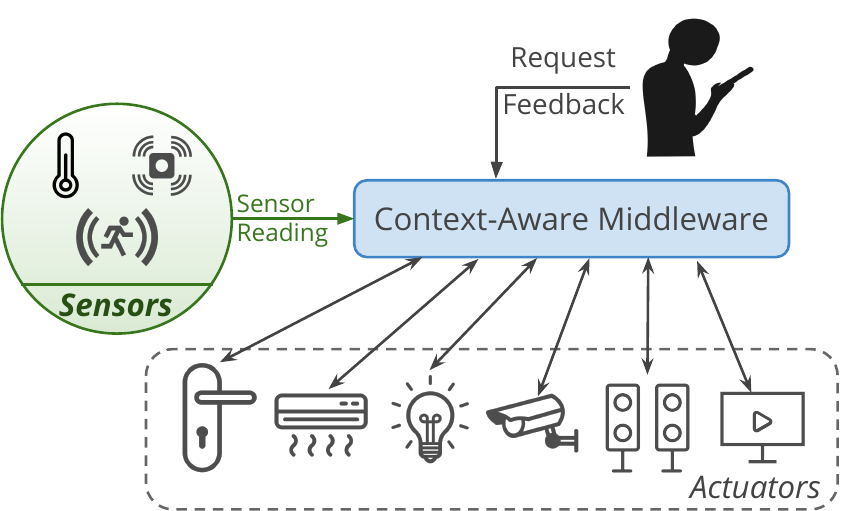}
%https://docs.google.com/drawings/d/14a5XR1fkuedKOJOAqN-FDyLHS_ux-H8oC1lQqu5Zx2Q/edit
\caption{Middleware oriented}
\label{fig:iotIndividual}
\end{subfigure}
\caption{Existing human-device interaction in the IoT}
\label{fig:iotExisting}
\end{figure}
{\bf Existing Middleware Architectures for the IoT.}
Fig.~\ref{fig:iotExisting} captures the architectures of the two primary control options available to Alice, given today's current technologies. The first, in Fig.~\ref{fig:iotManufacture}, is a {\em manufacturer-oriented} view in which Alice controls actuators through different manufacturer gateways and their (proprietary) applications. The obvious advantage is that manufacturers can provide comprehensive services for their devices. For Alice, this means she can enjoy the features that achieve ideal sound quality from the stereo system. On the other hand, Alice has to navigate steep and diverse learning curves associated with each of the manufacturers. Although some manufacturers allow users to define personalized automation based on primitive context information like time, they cannot leverage sensor data provided by other manufacturers~\cite{chew2017smart}, and as a result, they fail to fully respond to a user's more subtle intentions due to lack of context~\cite{yang2013learning}. In other words, these systems' approaches to ``personalization'' do not truly reflect the user.

Fig.~\ref{fig:iotIndividual} shows another option in which Alice can employ a general-purpose IoT middleware as a sort of universal gateway (e.g., IFTTT~\cite{ovadia2014automate}, Hydra~\cite{eisenhauer2010hydra}). The advantage is that Alice only needs to learn one control language that can also leverage contextual data collected by diverse sensors. The disadvantage is that Alice has to define all the interaction patterns by herself using some script language defined by the middleware. Even with current context-aware automation solutions~\cite{capra2003carisma, chahuara2017context, meng2016rule}, since the control interfaces are designed by a third-party, some device features  may not be supported. For example, it may not be feasible for a third-party framework to coordinate multiple speakers to provide manufacture designed sound effect. %\tocheckJH{In the next section, we will discuss existing context-aware solutions and what makes them less desirable for Alice.}

{\bf Context-Awareness in the IoT.}
An obvious pain point is the inability of existing middleware to internalize an expressive and complete notion of {\em context}, a need that has been identified in both the research community~\cite{sezer2018context} and in the industry~\cite{da2014internet}. 
% \jh{I think we can move this paragraph to the introduction}\tocheckJH{\sout{Given the recent advancement in mobile and IoT devices, it is entirely conceivable for applications to receive detailed, continuous, and real-time contextual data about a user~\cite{whitmore2015internet}, which in turn enables more expressive context-aware applications. On the other hand, it is not feasible for an individual user to individually connect to and control the available devices in a one-by-one manner. For instance, to enable intuitive and seamless interaction in the IoT, Alice's instructions should be as simple as ``turn the light on'' rather than ``set the light over the stove to its brightest level''.  Here, instead, context can be very informative; exactly what action(s) to take on what device(s) may depend on a variety of contextual factors, including Alice's location in her home, the presence (or not) of others in the space, the time of day, the level of ambient light, Alice's current activity, etc.}}
Existing work incorporating context-awareness into IoT-like applications adopts a semantic approach~\cite{kao2012user,maarala2017semantic,sun2015efficient,wei2013campus}, where context-awareness relies on a pre-defined ontology characterizing devices, users, and their relationships. In contrast, providing users seamless experiences requires an approach that does not rely on a user having {\em a priori} knowledge about how IoT devices affect the space in which they are located. This is necessary to ensure the approach is suitable for new spaces a user encounters for the first time or for spaces in which the devices or environment are dynamic.

CA4IOT~\cite{perera2014context} is a context-aware architecture that  selects sensors to provide context based on a {\em likelihood} index that captures the weighted Euclidean distance from the user's context and the context of the sensors. However, context reasoning is based only on a pre-defined static distance function with fixed contextual inputs. Probabilistic and association based solutions~\cite{nath2012ace,luxey2018sprinkler} provide efficient activity sensing and fluid device interaction, while other approaches use Hidden Markov Models (HMMs) to model context-awareness~\cite{chahuara2017context,chen2016context,mannini2010machine,ruvzic2014context}. These approaches either require a list of pre-defined ``situations'' to which they are restricted, or they make assumptions restricting the context and the environment. More recently, deep learning has provided context-aware activity recognition, interaction prediction, and smart space automation~\cite{hammerla2016deep, ordonez2016deep, radu2018multimodal}. Despite their promise, these approaches require very large data sets for training, which makes them not suitable for personalized approaches in which data is small.

The aforementioned approaches inspire our work. We target a more general space with diverse devices that may dynamically change. The key challenges of \textit{interactive machines} in general~\cite{suchman2007human}, articulates a gap between what systems can sense about the context and the user's actual intentions. That is, no matter how many sensors we use to capture context, gaps will exist in the system's knowledge. Therefore, unlike existing solutions, we emphasize that user feedback should be explicitly included in the decision-making process.

{\bf System Support for \systemname.}
Efficiently collecting context has been well studied. Through multi-device cooperation, continuous monitoring systems like CoMon~\cite{lee2016comon+} and Remora~\cite{keally2013remora} enable context generated by sensors to be consumed by applications executing on nearby smartphones. Self-organized context neighborhoods~\cite{liu18:pinch, liu2019stacon} built using low-end sensors have negligible communication overhead. It is exactly because of the availability of these cost-effective continuous sensing systems that \systemname's vision of IoT personalization can seamlessly incorporate expressive context in an IoT enabled space.

We rely on existing solutions to provide connectivity among heterogeneous IoT devices. The web-of-things~\cite{ferreira2013iot} makes devices available as web services and thus accessible through a canonical interface. Lightweight solutions~\cite{warble} opportunistically discover surrounding devices and control them through users' personal devices. In this work, we focus on how to utilize context to better select and control these devices.

% \note{\textbf{IoT middleware} connectivity and interoperability in IoT.}

%\input{rethink_ConAware.tex}
\section{An Overview of \systemname}\label{sec:overview}
In this section, we overview \systemname's core contributions and define its underlying key concepts. We describe our algorithms in detail in the following section. Our work targets smart spaces that contain multiple rooms equipped with IoT devices. There may be one or more users sharing the space, however we assume that requests from different users are compatible with each other (e.g., we assume that two users never simultaneously request different actions on the same devices).

\begin{figure}[tb]
\centering
\includegraphics[width=\columnwidth]{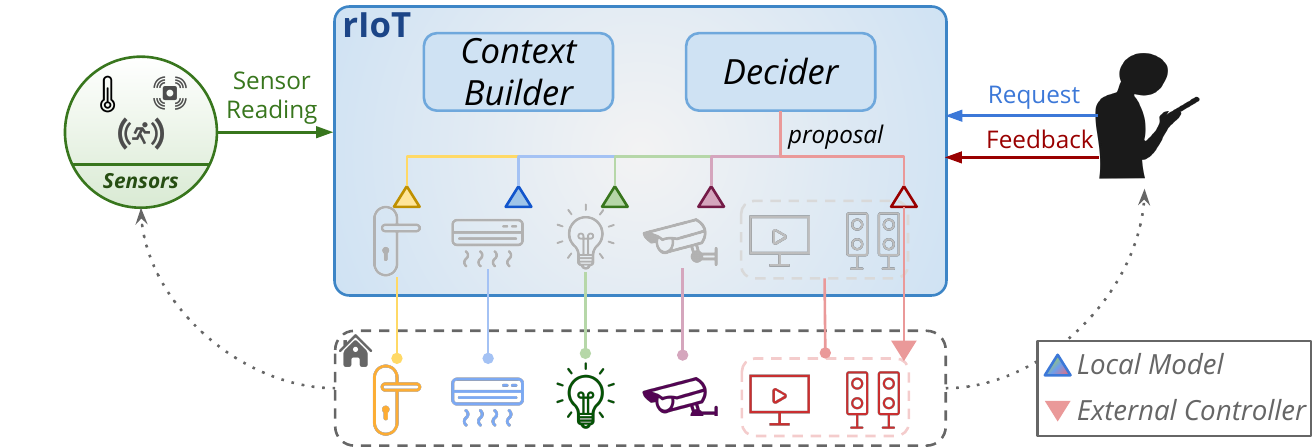}
%https://docs.google.com/drawings/d/1B4Yfg6N0I7v6NmCjZDl-HUpJLcn7zGcaQrYGeCRsiMI/edit
\caption{The overview of the \systemname framework}
\label{fig:iotParadigm}
\end{figure}
In Fig.~\ref{fig:iotExisting}, we identified a trade-off between user-oriented personalization and manufacturer-oriented features. We argue that it is important to enable personalization yet retain the full capabilities of devices. As shown in Fig.~\ref{fig:iotParadigm}, \systemname inserts itself between applications and IoT devices to allow applications to leverage context to automatically determine which devices and what actions on those devices best match a user's needs and expectations. \systemname encapsulates a {\em context builder} that collects and abstracts sensor readings into high-level, usable context. \systemname's {\em decider} uses context information, knowledge about available IoT devices, and knowledge about the user's prior interactions to choose (i) the best device  to fulfill a user's request and (ii) the best action to take on that device. 
%\systemname's contributions are focused on the components within the box labeled \systemname in Fig.~\ref{fig:iotParadigm}.

We assume that users' {\em requests} for IoT devices to take {\em actions} may be of varying levels of detail. At one end of the spectrum, a user may ask for a specific device to take a specific action (e.g., ``turn off the kitchen light''). At the other end, the user might simply ask for the IoT to ``act''. In this case, \systemname needs to determine which action on which device is most likely to satisfy the request. There are a variety of requests in between; for example, given the request ``turn on the light'', \systemname knows that the right action is to ``turn on'' and that the {\em type} of device is ``light'', but must determine which light device to act on. While we support all levels of specificity, in this paper, we focus primarily on the least specified, i.e., situations in which the user simply says ``act'', and \systemname must select the combination of device and action that best satisfy the user. 

\systemname learns a {\em local utility model} for each IoT device; this model ($f_d$) captures the likelihood that a given action on that device is the ``best'' action to take given a snapshot of the context at the moment that the user makes a request. Conceptually, each device proposes the action on that device that has the highest utility in the given context. \systemname's {\em decider} compiles all of the devices' proposals and selects the one with the highest overall utility. Given \systemname's choice of action, each device receives {\em implicit} feedback (i.e., if the device's proposal was selected, the device receives positive feedback; otherwise, the device receives negative feedback). Thus a device can learn about the utility of its actions in the context of other co-located devices. In addition, once the action is taken, \systemname allows the user to provide {\em explicit} feedback to reinforce (either positively or negatively) \systemname's selection.  The feedback is incorporated into the device's utility model, allowing it to learn over time based on the user's interactions in the space.

\systemname's architecture also allows applications to maintain access to manufacturer-specific actions. In such situations, \systemname controls the devices as a {\em system} through an {\em external} controller as  depicted to the right of Fig.~\ref{fig:iotParadigm}. Rather than the individual devices proposing an action, this controller proposes for all devices it controls. This allows exposing manufacturer-specific actions as part of the \systemname decision framework.

\noindent
We next define some terms that we use throughout the paper.
% \begin{defn}
% (user) In this paper, We assume that a user interacts with the IoT world with intent by issuing discrete requests. For each issued request, the user has an expectation of how the state of the IoT devices will change (e.g., that a light will come on or that the speaker volume will increase). The user will action taken indicating whether the system's r action taken indicating whether the system's r one \device doing one \action. We assume that the user will give one binary feedback for every action taken indicating whether the system's response was appropriate or not.
% \end{defn}
\begin{defn}\label{def:context}
(\context \underline{$c$}) 
%We follow the definition given in~\cite{dey2001understanding}, ``any information that can be used to characterize the situation of an entity''. 
Practically, a context $c$ is any single piece of numerical or categorical data and can be raw sensor reading like \textit{temperature} or \textit{illumination level}, or an abstract value derived from raw data, e.g., \textit{isAtHome}, \textit{Cooking}.
\end{defn}

\begin{defn}
(\context snapshot \underline{$C_t$}) We define $C_t$ as a vector of \context values $c_{t,i}$ that describe the user's situation at time $t$, i.e., $C_t=(c_{t,0}, \ldots, c_{t,i},\ldots, c_{t,n})$. We assume that the $i^{\it th}$ element of any snapshot is always the same type of context; $c_0$ is always the user's identity. 
% As a shorthand, we use the notation $c_{t,i}$ to refer just to the $i^{\it th}$ element of the context snapshot $C_t$.
% We assume the size and order of instances\cj{This is not clear. I think you mean that the universe of potential context labels is fixed and known {\em a priori}. I think what you want to say is more like ``'' We could add a footnote that this is achievable even if not all context types are used at all times by using name-value pairs.} in \context snapshot will remain the same. 
\end{defn}

% \begin{defn}
% (Location \context)\cj{This doesn't seem like it needs its own full-fledged definition. Instead, I'd follow up the terminology section with an assumptions section and include this there.} In this work, we assume that there exists a coordinate system for localization with some(small) error. Therefore, we define Location \context as the (x,y) coordinate of some location. 
% \end{defn}

% \begin{defn}
% (Time \context)\cj{This would go in assumptions, too, IMO. In this assumptions section, I would just say that, for the purposes of this paper, we'll assume the availability of some specific types of context: location, time, and activity. And then provide any constraints/notes about each of these.} In this work, we define the Time context as the second of the day. \systemname can support other time-based context but they are not discussed in this paper (i.e., dayOfWeek, dayOfMonth, etc.)
% \end{defn}

% \begin{defn}
% (Activity \context) The activity \context is represented as a string that describes the user's activity. We do not define a universal range of activities that a user can take but we assume that they will be consistent\cj{what does this mean?} for each user.
% \end{defn}

\begin{defn}
(\device \underline{$d$}) A \device is an actuator that can be discovered and controlled through a virtual controller.
\end{defn}

\begin{defn}
(\device class \underline{$T$}) A class $T$ is a set of \device{}s that have the same type, and therefore the same action interface, e.g., $d \in T_{\textit{light}}$. We assume a hierarchy of classes. For example, a dimmable light is also a light, i.e., $T_{\textit{dimmable}} \subset T_{\textit{light}}$. 
% If a user reque, ion \underline{$a$}) \tocheckCJ{\sout{We define a} A}n \action $a$ \tocheckCJ{\sout{as an} is some activity \sout{action}} that can be performed by an actuator\tocheckCJ{, e.g., \sout{like}} \textit{turnOn}, \textit{turnOff}, etc.
\end{defn}

\begin{defn}
(\action \underline{$a$}) An \action is performed by a device, e.g., \textit{turnOn}, \textit{turnOff}, etc. \underline{$A$} is the set of all actions; $A_d$ is the set of \action{}s \device $d$ can perform. We assume $A_d$ is finite.
\end{defn}
\begin{defn}
(\request \underline{$R$}) A request $R_t$ made by the user at time $t$ is a pair of class and \action, both of which are optional. Specifically, $R_t=\langle T, a\rangle$ indicates that the user wants a \device $d \in T$ to do action $a$. 
%A \request $R_t=\langle T, a\rangle$ is valid if $\exists d \in T \text{ , s.t. } a \in A_d$. 
A request's fields can be blank, i.e., $R=\langle \bot, \bot \rangle$, which indicates that the user requests the IoT to ``act'', or have only one of the two fields, e.g., $R=\langle light, \bot\rangle$ indicates a request for some light to take some action.
\end{defn}

% \begin{defn}
% (\proposal \underline{$Q$}) A \proposal is a tuple of \device, \action and a \utility. For example, $Q=\langle \textit{Light\_a}, \textit{turnOn}, 0.5 \rangle$ indicates that the utility value of 0.5 is associated with the action of turning on the device named \textit{Light\_a} at this moment. The \utility value will be used as part of our context learning algorithm.
% \end{defn}

\noindent
Formally, our problem statement is:
\begin{quote}
{\it Given a user's request $R_t$ at time $t$ and a snapshot of the context $C_t$ at the same time $t$, output a tuple $\langle d, a \rangle$ that specifies the action $a$ to be taken on device $d$ to best satisfy the request $R_t$}
\end{quote}

%\textbf{Problem formulation} The problem can be formulated as, when a user issues a request $R_t$ at time $t$, given the user's context snapshot $C_t$, \systemname should output a tuple $\langle d, a \rangle$ where $d$ is the device that should perform action $a$ to satisfy request $R_t$. 

\section{Context-Aware Decision Models in \systemname}\label{sec:riot_framework}
We now describe the components and processes that allow us to fill in the architecture in Fig.~\ref{fig:iotParadigm}. We then describe \systemname's contextually-driven learning algorithm in detail.

\subsection{\systemname Approach}\label{sec:sys_framework}
As described in the overview, \systemname learns a local utility model for each IoT device. Conceptually, these models ``belong'' to the devices themselves, but, as Fig.~\ref{fig:iotParadigm} shows, the models are part of \systemname. The only exceptions are {\em external} models, which may contain manufacturer-proprietary information; in these cases, the \systemname local model is a proxy for the external model, which resides under the manufacturer's purview.

\begin{figure}[t]
\centering
\includegraphics[width=\columnwidth]{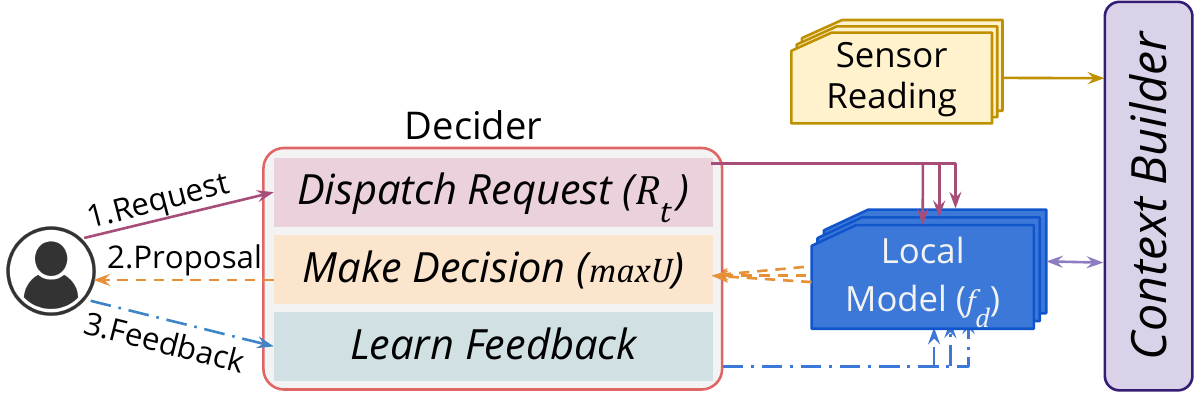}
\caption{The overview of data flow in \systemname }
\label{fig:system_framework}
\end{figure}
% \note{Change the line and ignore the external model, remove context registry}

Fig.~\ref{fig:system_framework} shows the flow of requests, context, and decisions in \systemname. The user (at the left) makes requests to \systemname's {\em decider}, which resolves them using input from the local utility models. These models in turn rely on context snapshots generated by the {\em context builder}. Given a decision, the user may accept the proposal (providing implicit positive feedback) or reject it (providing explicit negative feedback). This information is used to update the local models. In the case of negative feedback, the {\em decider} repeats the decision process and makes a new proposal to satisfy the original request.

\subsubsection{Building Context Snapshots}
The {\em context builder} translates raw sensor data into contextual data, which the local utility models use for learning. We rely on four generic \context abstractions: (1)~the {\em numeric context} allows the {\em context builder} to capture standard numerical values, e.g., temperature, pressure; (2)~the {\em cyclic numeric context} captures context types that are numerical but ``roll over'' on some predictable schedule, e.g., time, day of the week; (3)~the {\em N-dimensional vector context} captures context values that are represented by a tuple of values, e.g., location coordinates; 
% (4)~the {\em angular context}\cj{This one seems much more specialized than the others. Is this not just a cyclic context?} represents speed, direction, etc.; 
and (4)~the {\em categorical context} captures labeled values, e.g., human activity, binary data. Depending on the type of context and the available sensors, the {\em context builder} assembles the higher level values out of the raw sensor data. \systemname leverages existing work in context construction to implement the {\em context builder}.

\subsubsection{Context Distance Functions}
The devices' local utility models will propose actions to take in a given context based on the feedback they have received about prior actions in the same or similar contexts. To judge the similarity of two contexts, \systemname relies on context distance functions. We first define ${\it dist}(c, c^{\prime})$, or the distance between two contexts (e.g., the distance between two locations, the distance between two temperature values, etc.). Primitive context types typically have easily defined distance function (e.g., geometric distance, absolute value, cyclical distance). \systemname makes one additional constraint on any ${\it dist}(c, c^{\prime})$, i.e., that the distance is normalized to the range $[0,1]$. With this simple definition of contextual distance, we build a distance function ${\it dist}(C_a, C_b, W)$ that captures the distance between two context snapshots.

\begin{defn}
(context snapshot distance \underline{${\it dist}(C_a, C_b, W)$)} This distance is computed using the Manhattan distance~\cite{krause1986taxicab}; the vector $W$ weights the elemental values of the snapshots:
\begin{equation*}
    \begin{gathered}
    {\it dist}(C_a, C_b, W) = \sum_{i=0}^n{w_i\times{\it dist}(c_{a,i},c_{b,i})}\\
\mbox{where, } C_a = (c_{a,0},c_{a,1},\ldots,c_{a,n}), 
C_b = (c_{b,0},c_{b,1},\ldots,c_{b,n}), \\
W=(w_0, w_1,\ldots,w_n), (0<w_i<1)
    \end{gathered}
\end{equation*}
\end{defn}
Because the weight vector $W$ is an input to the function, each local model can use a different distance function for context snapshots, enabling personalization. For instance, some users may have a strict daily routine based on the clock, in which case a difference in time means more for this user than other context types. An interaction with lights is more likely based on location, while requests for a remote camera may depend more on suspicious sounds or movements.

Because context values can be continuous, it can be useful to discretize context snapshots into buckets. When we do so, we need to ask whether a bucket ``contains'' a context snapshot. We define ${\it contain}(c^l, c^u, c)$ for the first three elemental context types as $c^l \leq c \leq c^u$ and as $c \in (c^l \cup c^u)$ for categorical context. We next extend this to context snapshots:

\begin{defn}
(context snapshot contains \underline{${\it contain}(C_a, C_b, C_x)$)} This function simply requires the contain function for all of the elements of the snapshot to be true:
\begin{equation*}
    \begin{aligned}
    {\it contain}&(C_a, C_b, C_x)=\\
    &\begin{cases}
    \textsc{false}, \mbox{if } \exists i\mbox{, s.t. } {\it contain}(c_{ai}, c_{bi}, c_{xi})=\textsc{false}\\
    \textsc{true}, \mbox{otherwise}
    \end{cases}
    \end{aligned}
\end{equation*}
\end{defn}

\subsubsection{Defining and Using Local Utility Models}
\systemname's local utility models capture the suitability of devices' actions to requests from users in certain context states.
\begin{defn}
(local model \underline{$f_{d}$}) $f_d\colon C\times R \rightarrow A\times {\rm I\!R}$ maps a request and context snapshot onto an action the device can take and a {\em utility value}, $u\in [0,1]$. The utility captures the likelihood that the action is the ``right'' one given the request and context. $f_d(C,R)$ results in a proposal $P_d=\langle d, a, u\rangle$. 
\end{defn}

When a user requests an action, \systemname captures the context and requests a proposal from each device's local utility model. \systemname's objective is to output the final decision $P_{R_t,C_t} = \langle d,a,u\rangle$, which is the winning \proposal{}, i.e., the proposal with the maximum utility across all of the devices' proposals.

\systemname's key challenge is therefore how to compute the $f_{d}$ models for each device $d$. Because we do not want to make any assumptions about or place any constraints on the environment in which \systemname operates, our approach is to fit the $f_{d}$ models using each user's history of interactions with each device, leveraging similarities among the contexts of the interactions.

\subsection{Context Learning in \systemname}\label{sec:sys_algo}
%We now describe how \systemname learns local utility models ($f_d$) for each IoT device. Throughout, we demonstrate how \systemname works using an extension of our previous scenario.

Imagine that Alice has four IoT lighting devices in her home. Depending on the context, a ``turn light on'' request may indicate a desire to turn on any one of these lights. For instance, when Alice awakens at 6:30am and says ``turn light on'', she intends to illuminate the bedroom. While cooking in the kitchen (regardless of the time of day), a ``turn light on'' request should control the kitchen light. And when Alice is reading anywhere in her home, ``turn light on'' should affect the light closest to her current location. Alice's context snapshots may contain, for example, time (captured as a cyclic numeric context), Alice's location (captured, in a coordinate system), and activity (categorical context).

%\note{The user story is too long, and have some unneeded part.}
%\textbf{User Story.} Alice has four rooms: bedroom, bathroom, living room and kitchen. In the morning Alice wakes up at round 6:30 am, and she says "turn on the light" in the bed, which is a request to turn on the ceiling light($d_1$) in the bedroom. Alice cooks three meals a day. When she cooks in the kitchen, she will request the kitchen light($d_2$) to be on. During the day, Alice will read books in the living room and she will request the closest light from the two lights($d_3, d_4$) in the living room to be on. In the evening, before going to sleep Alice reads some books in the bed and says "turn on the light", which means the bedside lamp($d_5$) instead of the ceiling light($d_1$). 

% This story emphasizes the importance of contextual information(here location, time and activity) in smart space automation. Alice may request different lights in one location with different activity (i.e. waking up vs reading in the bedroom). She may also request different lights for one activity in different location (i.e. reading in the living room vs reading in the bedroom). 
%The \context types we will use in this example are \textit{Time}(Cyclic Numeric Context), \textit{Location}(1-d Coordinate Context), \textit{Activity}(Categorical Context). For example, we Alice is sleeping in bed at 11 pm, the context is $C=(Alice, 23:00,1,sleep)$ where 1 is a possible \textit{Location} coordinate of being in bed. The coordinate of the bed is (0-1), living room(3-6), kitchen(7-9).

The goal of \systemname is to learn each local utility function $f_{d}(C, R)$ based on the user's interactions and feedback. However, learning this function directly can be challenging. Context can be continuous, and the target function may be non-linear and non-continuous. Therefore, rather than learning $f_{d}(C, R)$ directly, we introduce a piecewise function set as an approximation, based on a concept we call {\em state}:
\begin{defn}
(state \underline{$S$}) A state is defined by three \context snapshots, $S=(C_{{\it min}}, C_{{\it max}}, C_{{\it mid}})$. We use two functions to compare states: ${\it contain}(S, C_x) \equiv {\it contain}(C_{{\it min}}, C_{{\it max}}, C_x)$, determines whether a state contains a given context snapshot, while ${\it dist}(S, C_x, W) \equiv {\it dist}(C_{{\it mid}}, C_x, W)$ computes how far a context snapshot is from the state.
%, where $W$ is the weight vector as introduced before. 
We define the {\em  radius} of a state as $r_S \equiv {\it dist}(C_{{\it max}}, C_{{\it min}})/2$.
\end{defn}
The set of states that discretizes the space of context snapshots is not defined {\em a priori} but are learned over time. The learned states need not cover the entire space of possible context snapshots, and different devices can have different relevant states. In our scenario, Alice's lights may learn states defined by ranges of time and activity labels; her living room lights may not learn anything about states in the very early morning, though her bedroom lights will.

% \begin{defn}
% (local utility of a state, \underline{$\hat{f}_{d,a}(S, R)$}) Each element of the piecewise approximation of $f_d$, is a function $\hat{f}_{d,a}\colon S\times R \rightarrow {\rm I\!R}$ that captures the utility of taking action $a$ on $d$ when the context is contained by $S$. 
% \end{defn}
% Given a request $R = \langle T, a_{\it req}\rangle$ in which $T$ is a (potentially empty) device type and $a_{\it req}$ is a (potentially empty) requested action, $\hat{f}_{d,a}(S, R)$ is undefined if $(T \neq \bot \wedge d\not\in T)$ or $(a_{\it req}\neq\bot\wedge a\neq a_{\it req})$. To bootstrap the learning of the local utility models, we initialize the local utility of a state to a non-zero default value. In particular, in this paper, we use the following initialization:
% \begin{equation*}
%     \begin{aligned}
%     \hat{f}_{d,a}(S, \langle T, a \rangle) =
%     \begin{cases} 
%     0.5 \text{   , if } (T = \bot\vee d \in T) \text{ and }\\
%     \hspace{1.1cm}(a_{\it req}=\bot\vee a=a_{\it req})\\
%     0 \text{  , otherwise}
%     \end{cases}
%     \end{aligned}
% \end{equation*}

We use this concept to approximate each device's local utility model by combining a utility learned for each state.
\begin{defn}
(local utility of a state, \underline{$\hat{f}_{d,a}(S)$}) Each function $\hat{f}_{d,a}=u\colon S \rightarrow {\rm I\!R}$  captures the utility of taking action $a$ on $d$ when the context is contained by $S$. The default value is 0.5 which means the likelihood action $a$ is a good choice is 50\%.
\end{defn}

Given a request $R = \langle T, a_{\it req}\rangle$ in which $T$ is a (potentially empty: $T=\bot$) device type and $a_{\it req}$ is a (potentially empty: $a_{\it req}=\bot$) requested action, device $d$'s local utility function $f_{d,a}(C, R)$ can then be approximated as:
\begin{equation*}
\begin{gathered}
 A_{R} = 
    \begin{cases}
        A_d  \text{, if } a_{\it req} = \bot\\
        {a_{\it req}} \text{, otherwise }
    \end{cases}\\
u_{\it max} = \langle {\rm max}\:a, S : {\it contain}(S, C) \wedge a\in A_{R} :: \hat{f}_{d,a}(S)\rangle\rangle\\
    \hspace{0.2cm}f_{d,a}(C, \langle T, a_{\it req} \rangle) =
    \begin{cases} 
    \langle a_{\it max}, u_{\it max}\rangle \text{,if} (T = \bot\vee d \in T)\wedge\\ 
        \hspace{1.8cm} (a_{\it req}=\bot\vee a_{\it req} \in A_d)\\

    \langle a_{\it req}, 0 \rangle \text{, otherwise}
    \end{cases}
    \end{gathered}
\end{equation*}

% \sout{$\{\hat{f}_{d,a_0}(S, R), \hat{f}_{d,a_1}(S, R)\ldots\}$ to approximate $f_{d}$. Each function $\hat{f}_{d,a}\colon S\times R \rightarrow {\rm I\!R}$ is defined as: $\hat{f}_{d,a}(S, R) = u$, where $u$ is the utility associate with $d$ doing $a$. And $f_{d}(C, R)$ can computed by $f_{d}(C, R)= \langle a_{max}, \max_{a\in A_d}\hat{f}_{d,a}(S, R) \rangle$, if $conatin(S,C)=ture$, where $S$ and $contain(S,C)$ are defined as:}}

% The learning of functions $f_d$ with infinite points is approximated to the learning piecewise functions $\hat{f_d}$ with finite sections, which can be stored as arrays.
% To bootstrap the learning of the local utility models, we initialize the local utility of a state to a non-zero default value. In particular, in this paper, we use the following initialization (where the request $R_t$ is contains a (potentially empty) device type $T_t$ and a (potentially empty) requested action $a_t$):
%\begin{equation}
%    \begin{aligned}
%    \hat{f}_{d,a}(S, \langle T_t, a_t \rangle) =
%    \begin{cases} 
%    0.5 \text{   , if } d \in T_t \text{ and } a=a_t\\
%    0 \text{  , otherwise}
%    \end{cases}
%    \end{aligned}
%\end{equation}

\begin{algorithm}[tb]
\caption{Computing the Local Model}
\label{alg:req}
\DontPrintSemicolon
{\small
\SetKwProg{onReceiveRequest}{Function {\sc onReceiveRequest:}}{}{end}
\SetKwProg{onFeedback}{Function {\sc onFeedback:}}{}{end}
${\bf S}$: set of known states, initially empty\\
$f_d(C,R)$: set of piecewise local functions, initially empty\\
    \onReceiveRequest{$R=\langle T, a\rangle$, $C_t$}{
        \If{($T \neq \bot\wedge d \notin T$) or ($a_{\it req}\neq\bot\wedge a_{\it req} \notin A_d$)}
        {return $\langle a_{\it req}, 0 \rangle$}
        \leIf{$a = \bot$}
            {let ${A_{R}} \leftarrow A_d$}
            {let ${A_{R}} \leftarrow \{a\}$}
        \If{$\not\exists S_i \in {\bf S}$, {\rm s.t.} ${\it contain}(S_{i}, C_{t}) = $ {\sc true}}{
        $S_{\it new} \leftarrow (C_t - r, C_t + r, C_t)$\label{line:radius}\\
        \lIf{${\bf S} = \emptyset$}{
            $\forall a\in A_d$, $\hat{f}_{d,a}(S_{\it new}) \leftarrow 0.5$\label{line:defaultutility}
        }
        \lElse{$\forall a\in A_d$, initialize $\hat{f}_{d,a}(S_{\it new})$ from neighborhood}\label{line:neighborhood}
        ${\bf S}\gets {\bf S}\cup \{ S_{\it new}\}$\\
        $f_d(C,R)\gets f_d(C,R)\cup \{\hat{f}_{d,a}(S_{\it new})$ : $ a\in A_d\} $
        }
        let ${\bf S}_{R} \leftarrow \{S_i : {\it contain}(S_{i}, C_{t}) = $ {\sc true}$\}$\label{line:startmax}\\
        let $u_{\it max} \leftarrow \langle {\rm max}_{a\in A_R, S_i\in {\bf S}_{R}}: \hat{f}_{d, a}(S_i)\rangle$\\
        let $a_{\it max} \leftarrow a$ s.t. $\hat{f}_{d, a}(S_i) = u_{\it max}$\\
        {\bf return} $P_d = \langle d, a_{\it max}, u_{\it max}\rangle$\label{line:endmax}\\
    }
    \onFeedback{$P=\langle d, a, u\rangle$, $C_t$, ${\it feedback}$}{
        \For{$S_{i} \in {\bf S}$  {\rm s.t.} ${\it contain}(S_{i}, C_{t})=$ {\sc true}}{
             \eIf{${\it feedback}$ {\rm is} positive}{
                $\hat{f}_{d, a}(S_{i}) \gets \hat{f}_{d, a}(S_{i}) + {\it reward}$    
            }        
            {
                $\hat{f}_{d, a}(S_{i}) \gets \hat{f}_{d, a}(S_{i}) - {\it reward}$             
            }
        }
    }
}
\end{algorithm}

Algorithm~\ref{alg:req} shows our approach. When receiving a request $R=\langle T, a \rangle$, if $C_t$ is not ``contained'' in any states, \systemname will create a state with a default radius $r$ around $C_t$ (Line~\ref{line:radius}). This new state's utility is the default (Line~\ref{line:defaultutility}) or initialized based on other ``nearby'' states (Line~\ref{line:neighborhood}; explained in detail below). Once $C_t$ is ``contained'' in a known state, \systemname will output the action that has the highest \utility of all actions and is compatible with the request $R$ (Lines~\ref{line:startmax}-\ref{line:endmax}). When receiving feedback from the user, \systemname takes the prior \proposal $P$ and the prior \context, $C_t$, and updates the local model. The ${\it reward}$ is computed using the {\em Sigmoid} function; we translate the utility from $(0,1)$ to $(-\infty, +\infty)$, increment or decrement it by a constant $reward$, and then translate it back to $(0,1)$. This adjusts the utility more slowly when it is close to 0 or 1.

% We use some interaction examples in Alice's scenario to show the issues and how we improve the primitive approach.
\subsubsection{Initializing Models from Nearby States}~\label{sec:neighbor}
% In the primitive algorithm, the \utility values are all initialized to 0.5. However, once we have some interactions with the user, a new state should be able to learn knowledge from the previous interactions. In this user example:
% Alice always wakes up at 6:30 and request the bedroom ceiling light($d_1$) on. The local model is going to learn that at the state where she is around her bed and the time is around 6:30. The \utility for turning on $d_1$ is very high. It can be shown as: $R_{1}=\langle \textit{Light}, \text{turn on}\rangle$, $C_1=(Alice, 6:30, 1, wakeUp)$(Note that 0-1 is the possible \textit{Location} of Alice being in bed). The \utility for $\hat{f}_{d_1,a_1}(S_1, R_1) \approx 1$ where $ a_1 = \textit{turnOn}, conatin(S_1, C_1) = \text{true}$. 
% One day, Alice is sick and she wakes up at 9:00 instead of 6:30. She issues the same request as usual $R_{2}=\langle \textit{Light}, \text{turn on}\rangle$ and expect the $d_1$ is turned on. Although this is the first time(9:00) she issues the request, which means no model is learned $\hat{f}_{d_{1},a_2}(S_2, R_1) =0.5$ where $ a_2 = \textit{turnOff}, conatin(S_2, C_2) = \text{true}$, we do not want the algorithm give a random decision. Instead, since Alice always request the ceiling light in bed at 6:30, it should be very likely that she also wants the ceiling light this time. To realize this idea, we employ a weighted nearest neighbor method to initialize the unlearned part of the model in an unsupervised way.
%Like many distance based approaches, 
We assume that a user will have similar behaviors in similar contexts~\cite{perera2014context}. Therefore, when initializing a new state $S$, \systemname computes the initial utility values based on utility values for nearby states. For example, if Alice's actions routinely trigger the bedroom light at 6:30am, even the first time she requests a light at 7:00am, it is likely that she also wants the bedroom light. To capture this ``nearness'' in \systemname, the first time a user makes a request in a new state, we use the $k$ learned states that are closest to the new \context and use their learned models as a basis for the new state's utility value; a state's contribution is weighted by its distance to the new context. More formally:
\begin{equation*}
\label{eq:kNN}
    \begin{aligned}
\hat{f}_{d,a}(S_{\it new}) = 0.5+\frac{1}{k}\sum_{S_{i} \in \textbf{S}_{\it kNN}} \frac{\hat{f}_{d,a}(S_{i}) - 0.5}{dist(S_i, C_{\it new}, W_{d})/r_{S_{\it new}}+ 1}
    \end{aligned}
\end{equation*}
Since 0.5 is the default value, we first shift the utility value to it and weight it with the distance plus one to make sure that the derived utility value is closer to 0.5 than the original value to prevent initializing the new state with too much confidence.

Although the assumption that behaviors in ``nearby'' states are likely to be similar is valid in general, this assumption can sometimes be misleading, for example, when there is a wall between two locations. In such a case, the \utility value we ``borrow'' from the neighborhood can be very wrong. Therefore, we add a flag to any new state $S_{\it new}$. If the user gives negative feedback immediately following initialization, we abandon the initialized \utility value and reset it to 0.5.

\subsubsection{Decision-Tree Like State Splitting}~\label{sec:entropy}
\systemname {\em \ splits} states as its models learn more nuanced behaviors of users. We might initially learn the difference in behaviors between morning and afternoon; over time, differences between early morning and late morning might become apparent. Our approach is inspired by the splitting used in the ID3 decision tree algorithm~\cite{quinlan1986induction}. 
%In the previous algorithm, the state will never be changed after it is created. In practice, this could be problematic. If a state is at the decision boundary, its feedback may be half positive and half negative. For example, if Alice requests the bedroom light from 6:00-7:00 and living room light from 7:00-8:00. If the first request Alice made was at 6:50, so a state $S_p$ is created that contains the context from 6:30 to 7:30, more formally, $contain(S_{p}, (Alice, 6:30-7:30))=true$. Now $S_p$ is a uncertain state because there is no correct light for this state. 
To detect candidate states for splitting, we define {\em entropy} using the feedback the device has received for actions taken in each state. Conceptually, entropy captures how internally different the feedback is for a given state. If a state has an even mix of positive and negative feedback, it gives ``wishy-washy'' proposals; it therefore makes sense to look for a way to split the state into two that are more internally consistent. 
\begin{defn}
(feedback cache \underline{$\Phi_S$}) For each state, a device stores a feedback cache $\Phi_S = \{ (C_1, P_1, b_1), (C_2, P_2, b_2),\ldots\}$. Each element $\phi_i\in\Phi_S$ contains the context in which an action was recommended ($C_i$), the proposal that was made by the device ($P_i$), and the (boolean) feedback ($b_i$).
\end{defn}
%We first detect these uncertain states using a metric called entropy, the definition of entropy is derived from the decision-tree algorithm with the similar idea:
\begin{defn}
(entropy \underline{$E_{S}$}) %For each state, we store a feedback cache $\{(C_0, Q_0, b_0), (C_1, Q_1, b_1)\ldots\}$, which is a set of request/feedback history. Each element $shot_i$ in the cache is a snapshot containing a \context $C_i$ in which the user make a \request, a proposal $P_i=\langle d_i, a_i, u_i\rangle$, which is the proposal made by device $d_i$, and a feedback $b_i$, which is the feedback given by the user. $b_i =0$ means negative feedback and 1 means positive feedback. Now we define the Entropy of a state $S$ as:
To compute entropy, we first compute the fraction of positive (and negative) feedback for each action $a\in A_d$:
\begin{equation*}
    p_{\it pos}(S, a) = \frac{|\{\phi: \phi\in\Phi_S \wedge \phi(a) = a \wedge \phi(b) = 1\}|}{|\{\phi : \phi\in\Phi_S \wedge \phi(a) = a\}|}
\end{equation*}
We use $\phi(a)$ to indicate the action associated with the proposal in $\phi$ and $\phi(b)$ to indicate the boolean feedback associated with $\phi$. The value of $p_{\it neg}(S, a)$ is defined similarly but constraints $\phi(b)$ to be 0. We then compute the entropy of state $S$ as:
\begin{align*}
E_{S}=\max\limits_{{a\in A_d}}\{&-p_{\it pos}(S, a)\log(p_{\it pos}(S, a))\\ &- p_{\it neg}(S, a)\log(p_{\it neg}(S, a))\}
%\text{where, } &p_{pos}(S,a)= \frac{\text{the number of }shot_{i} \in \text{Cache}_{S}\textbf{, s.t. } a_i = a, b_i=1}{\text{the number of }shot_{i} \in \text{Cache}_{S} \textbf{, s.t. } a_{i}=a}\\
%&p_{neg} \text{ is defined the same as } p_{pos} \text{ but with } b_{i} =0
\end{align*}
%\label{def:entropy}
Intuitively, the more similar $p_{\it pos}$ and $p_{\it neg}$ are, the more ``indecisive'' the state is, and the higher the entropy value.
\end{defn}

%\begin{algorithm}[tb]
%\caption{State Splitting}
%\label{alg:split}
%{\small
%\SetKwProg{onFeedback}{Function {\sc onFeedback:}}{}{end}
%        \onFeedback{$P=\langle d, a, u\rangle$, $C_{t}$, ${\it feedback}$}{
%        \nonl$\ldots$ as before $\ldots$\\
%        \For{$S_t$ \text{from }\textbf{S} \text{ s.t. }$contain(S_t, C_t)=$true}{
%            add $(C_{t}, P_{t}, feedback)$ to $\text{Cache}_{S_{t}}$\\
%            \If{$E_{S_{t}} > \textit{threshold}$}{
%                $\textit{maxGain} \gets \displaystyle \max_{S_{1},S_{2}}(E_{S_{t}}-E_{S_{1}}-E_{S_{2}})$\\
%                \eIf{$\textit{maxGain} > \textit{minialGain}$}{
%                    Split $S_t$ into $S_1$, $S_2$ 
%                }{
%                    NOP
%                }
%            }
%         }
%    }
%}
%\end{algorithm}

\begin{algorithm}[tb]
\caption{State Splitting}
\label{alg:split}
{\small
\SetKwProg{onFeedback}{Function {\sc onFeedback:}}{}{end}
        \setcounter{AlgoLine}{19}
        \onFeedback{$P=\langle d, a, u\rangle$, $C_{t}$, ${\it feedback}$}{
        \For{$S_{i} \in {\bf S}$  {\rm s.t.} ${\it contain}(S_{i}, C_{t})=$ {\sc true}}{
            %\eIf{${\it feedback}$ {\rm is} positive}{
            %    $\hat{f}_{d, a}(S_{i}) \gets \hat{f}_{d, a}(S_{i}) + {\it reward}$    
            %}        
            %{
            %    $\hat{f}_{d, a}(S_{i}) \gets \hat{f}_{d, a}(S_{i}) - {\it reward}$             
            %}
            \nonl$\ldots$ {\it as in} Algorithm~\ref{alg:req}$\ldots$\\
            \setcounter{AlgoLine}{26}
            $\Phi \gets \Phi\cup \{(C_t, P, {\it feedback})\}$\\
            \If{$E_{S_{i}} > \textit{threshold}$}{
                $\Psi \gets \{(S_1, S_2)$ : all pairs of $S_1$ and $S_2$ split $S_i$ along every context value$\}$\label{line:split}\\
                $\textit{maxGain} \gets \displaystyle
                \max_{(S_{1},S_{2})\in\Psi}(E_{S_{i}}-(E_{S_{1}}+E_{S_{2}}))$\label{line:infogain}\\
                \If{$\textit{maxGain} > \textit{requiredGain}$}{\label{line:reqGain}
                    Split $S_i$ into $S_1$, $S_2$ 
                }
            }
         }
    }
}
\end{algorithm}

Algorithm~\ref{alg:split} shows how we extend the {\sc onFeedback} function to split states. After a new piece of feedback is incorporated into the local model and added to the feedback cache, the new entropy of the state is computed. If the entropy exceeds a \textit{threshold}, then \systemname determines whether a split of the state $S_i$ is preferable. Specifically, \systemname mimics the ID3 algorithm, using context as the attribute for splitting~\cite{quinlan1986induction}. In \systemname, each split will be based on a single context. We compute all pairs $(S_1, S_2)$ that split $S_i$ according to context value. For categorical context, we simply split along all possible categories. For continuous contexts, we use a constant number of split points for each context value, splitting  the range $(C_{\it min}, C_{\it max})$ evenly, given the target number of split points (Line~\ref{line:split}). For each pair $(S_1, S_2)$, we compute the information gain of splitting, which is defined as the entropy of the original state ($S_i$) minus the sum of the entropies of the two new states after splitting (Line~\ref{line:infogain}); we choose the split $(S_1, S_2)$ that has the maximal information gain (\textit{maxGain}) as the final candidate. To avoid overfitting, we use a second threshold, \textit{requiredGain}, as a required lower bound for information gain to achieve before we split (Line~\ref{line:reqGain}).

\section{Evaluation}\label{sec:evaluation}
In this section, we evaluate \systemname on two real-world data sets~\cite{cook2012casas}\footnote{The data sets (HH118, HH107) are at {http://casas.wsu.edu/datasets/}.}. We sought to answer the following questions:
\begin{enumerate}
\item How does \systemname perform on (noisy) real-world data?
\item How sensitive is \systemname to the specificity of context types?
\item How do request details assist \systemname in decision making?
\item How resilient is \systemname to dynamics?
\item Is it feasible for \systemname to run on mobile devices?
\end{enumerate}
In our experiments, we compare the performance of \systemname's algorithms to two machine learning approaches: random forest (RF)~\cite{ho1995random} and multi-layer perceptron (MLP)~\cite{pal1992multilayer}.

% \tocheckCJ{\sout{The main objective of the first experiment is to evaluate how \systemname performs in terms of accuracy in real world scenarios with noisy sensor readings. The second experiment does a sensitivity analysis of the accuracy for different \context{} types and \request{} types. In the third experiment, a virtual user is simulated based on the activities in a real-world data set. The purpose of this experiment is evaluate how \systemname is resilient to environmental change. In the first and third experiment, we use Random Forest(RF) and Multi-Layer Perceptron(MLP), which are well studied and suitable for tabular data, as the comparison group.}}

% In the experiments, we test different combinations of four types of context: Location, Time, Activity, and the raw motion sensor reading. Our choice of \context{}s is representative of the user's daily interaction, as shown in indicated in this simple user study~\cite{portet2013design}.
% In the first experiment, we use the raw data with no pre-processing and compare our result with 2 different machine learning algorithms. In the second experiment, we use the user's daily activities, which are labeled by the user, with simulated \context and more complex user scenarios. 
\subsection{\systemname in the Real World}\label{sec:exp_realworld}
In our first experiment, we use a data set (HH118) generated from seven months of interactions of an adult living in a smart home. The data set includes data from 18 infrared motion sensors which each generate a binary output indicating whether a human is present. From the data set, we derived context values from the raw sensor readings: 
\begin{itemize}
    \item {\bf current location} [categorical context] -- a label associated with the most recently activated motion sensor
    \item {\bf previous location} [categorical context] -- a label associated with the most recently de-activated motion sensor
    \item {\bf second of the day} [cyclic numeric context] -- the time
\end{itemize}
The home contained 10 light devices. We used the dataset's light on/off events to construct requests in \systemname. All such generated \request{}s are the least specified type $R=\langle {\it \bot}, {\it \bot} \rangle$, meaning the user simply requests the environment to ``act'' without providing a device type or action to take, however since there are only light devices in this experiment, the request is the same with or without device type. We compared \systemname with MLP, RF, and a baseline that always selects the light closest to the user's location. For ground truth, we used the light that was actually chosen by the user in the dataset.

Among the alternatives, only \systemname supports re-selection after negative feedback. Therefore, we use a metric, \textbf{First Decision Accuracy (FDA)} that evaluates the success of the first decision made by each approach; we define {\bf FDA} as \textit{the percentage of requests that are satisfied by the first decision}. 
% \begin{gather*}
% \text{First Decision Accuracy}(FDA)=\\
% \frac{\text{the number of requests statisfied by the first decision}}{\text{the number of requests}}
% \end{gather*}
\begin{figure}
\centering
\includegraphics[width=\columnwidth]{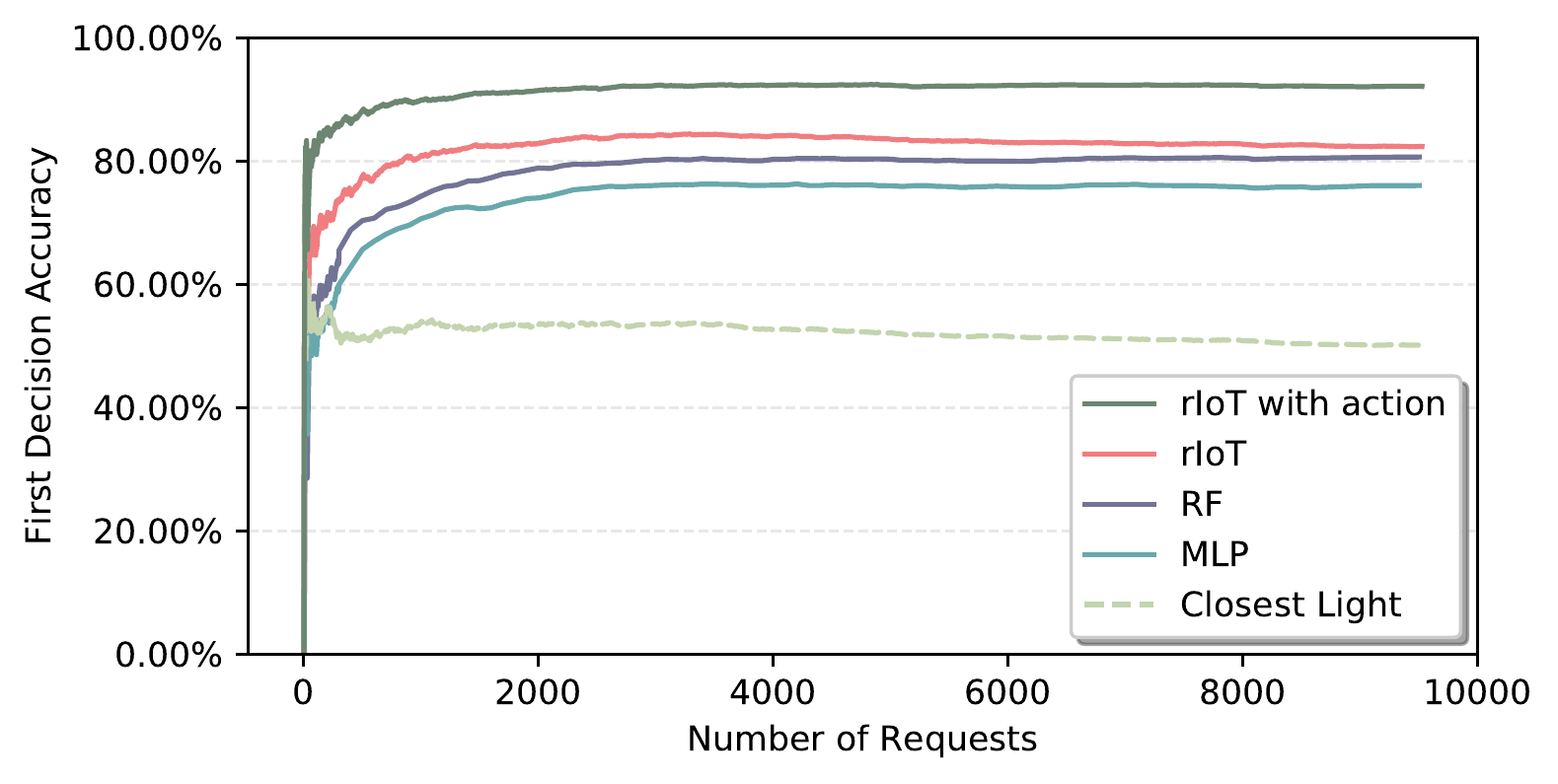}
\caption{{\it Experiments 1 and 2.} The figure shows the learning curves for each approach on the sequence of 9523 requests we derived from the data set. The curve ``\systemname with action'' is the result when the request states the action (``on'' or ``off'') to take. \systemname learns faster and has higher overall accuracy.}
\label{fig:real_compare}
\end{figure}

% \note{explain only light, the ``act'' is rally with device type. Keep the order of legend the same with lines. Maybe no formula. 0 to 100\% include the other request type in the same graph}
% For \systemname, we generate two categorical \context{}s from the motion sensor reading when receiving a request: ``which motion sensor is on most recently'' and `` which is off most recently''. The meaning of the two \context{}s are ``where is the user'' and ``where did the user come from''.
Fig.~\ref{fig:real_compare} shows the results. For \systemname, we compute the accumulated \textbf{FDA} at each \request. For RF and MLP, we re-train the model after every 100 requests. The hyper-parameters for RF and MLP are tuned during each training process. \systemname not only has the overall best \textbf{FDA} at 82.35\%, but it also learns significantly faster than the alternatives. \systemname requires 2059 fewer requests (about 50 days) than RF to reach 80\% accuracy and 863 fewer requests (about 20 days) than MLP to reach 70\%. In addition, \systemname provides immediate correction if the user gives negative feedback; in this experiment, 98.5\% of the \request{}s are satisfied by the first two decisions in \systemname.

%This experiment shows that \systemname can provide seamless human-device interaction in a real IoT environment with noisy sensor readings and no setup cost from the user. The benefit of using \systemname is that the user can just throw sensors and devices into an IoT space and use them immediately with no need for manual setup tasks.

% The key finding of this experiment is that \systemname works well with noisy sensor readings. Although the source of the data set does not mention the uncertainty in the response time of the motion sensors used, wide-area infrared motion sensors in general could have a delay time up to minutes\footnote{These are two commercial infrared motion sensors, the delay time are up to 5 and 10 minutes, respectively. \url{https://www.mpja.com/download/31227sc.pdf},\url{https://circuitdigest.com/electronic-circuits/pir-sensor-based-motion-detector-sensor-circuit}}. 

\subsection{\systemname's Sensitivity to Context and Request Detail}
We next experiment with different \context{}s and \request{}s to show that when the user provides more information, \systemname can opportunistically improve its performance.

\subsubsection{Request Specificity}
We first demonstrate how providing more detailed requests assists \systemname. In particular, when the user specifies the particular action to take (e.g., ``on'' or ``off'') \systemname's achieved {\bf FDA} improves (see Fig.~\ref{fig:real_compare}). With the specificity of the request, the overall \textbf{FDA} is 92.10\%, compared to 82.35\% when the specific action is not provided.

\subsubsection{Representing the User's Location}
Although motion sensor readings can categorically label the location of the user, this sacrifices the ability to compare locations, which in turn hinders \systemname's ability to leverage similarities between \context{}s. Therefore, we evaluate the performance of \systemname in the same setup as above but with location represented as a two-dimensional coordinate rather than a label.
%\footnote{We manually computed the coordinate of each motion sensor from the floor plan provided with the data set.}. 
We compute the \textbf{Average Feedback per Request (AFR)}, defined as {\it the average number of negative feedbacks the user gives per request} before \systemname selects the right device. 
\begin{figure}
\centering
\includegraphics[width=.9\columnwidth]{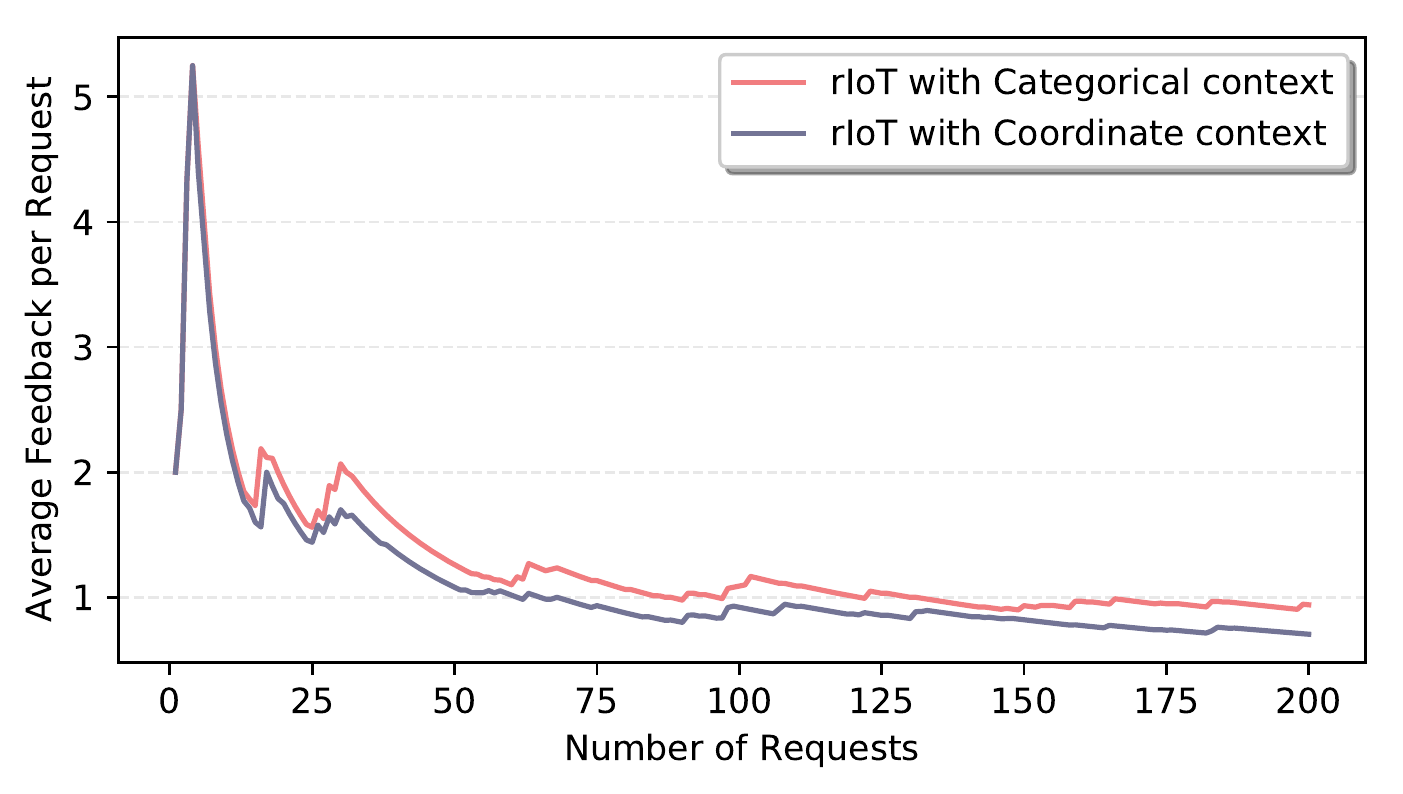}
\caption{{\it Experiment 3.} This figure shows the impact of context specificity on \systemname by contrasting the use of categorical location context versus coordinate context. The user needs to give fewer negative pieces of feedback for coordinate context because \systemname is able to learn from similar requests.}
\label{fig:real_context}
\end{figure}

Fig.~\ref{fig:real_context} shows the results for the first 200 requests. In the first several requests, the two options have the same \textbf{AFR} because \systemname has not yet learned from feedback. However, after 11 requests, \systemname with coordinate context initialized new states based on information from similar contexts. Quantitatively, \systemname with categorical context requires the user to give 39 more pieces of feedback in the first 200 requests, which increases the user's overhead and frustration in employing \systemname.

\subsection{\systemname's Resilience to Dynamics}\label{sec:exp_semi}
Our next goal was to determine how \systemname responds to changes in the IoT deployment. We generated simulated \context{}s and \request{}s so that we could inject uncertainty in sensor readings and changes in  IoT devices. We simulated a user interacting with lights, cameras, and speakers based on the real activities in our second dataset (HH117). We created 6393 requests from 26 different daily activities performed over one month by two real people; the labeled activities included sleeping, bathing, cooking, reading, working, watching TV, etc. We created ground truth of how each user interacts with devices during these activities. For example, we assumed that, when the user is engaged in the bathing activity, the user will turn on the bathroom light or when watching TV in the living room, the user will turn on the living room speaker. We similarly assigned each activity to a location context based on the particular activity. If an activity had multiple possible locations (i.e., reading happens both in the living room and bedroom), we used the motion sensor reading in the original dataset to decide which room is the true location for the instance of the activity. We built each context snapshot using the time of the activity and a random x-y coordinate sampled from the room in which the activity was occurring. To simulate habitual interaction patterns, we added three activity types:
\begin{itemize}
    \item \textbf{Wake up}: the {\em wake up} activity was added after every {\em sleep} activity. Our simulated user desired the bedroom speaker to play soft music upon waking.
    \item \textbf{Evening news}: the {\em evening news} activity was added every day the user was home at 18:00. Our simulated user desired to turn on the living room speakers no matter where he was located or what else he was doing. 
    \item \textbf{Doorbell}: we added a randomly generated {\em doorbell}; when this event occurred, our simulated user would desire the doorbell camera to be turned on.
\end{itemize}
All the \request{}s specified the device type that the user wants to use and related action, e.g. $R= \langle \textit{light}, \textit{turnOn} \rangle$; \systemname's task was therefore to determine the most appropriate device.

Before considering dynamics, we compared the overall \textbf{FDA} of the three approaches; the accuracies of \systemname, MLP, and RF were 98.92\%, 84.06\%, and 84.12\% respectively. With no artificial inaccuracies in the contextual data, \systemname very closely captures all of the users' interaction patterns. 

\begin{figure}
\centering
\includegraphics[width=\columnwidth]{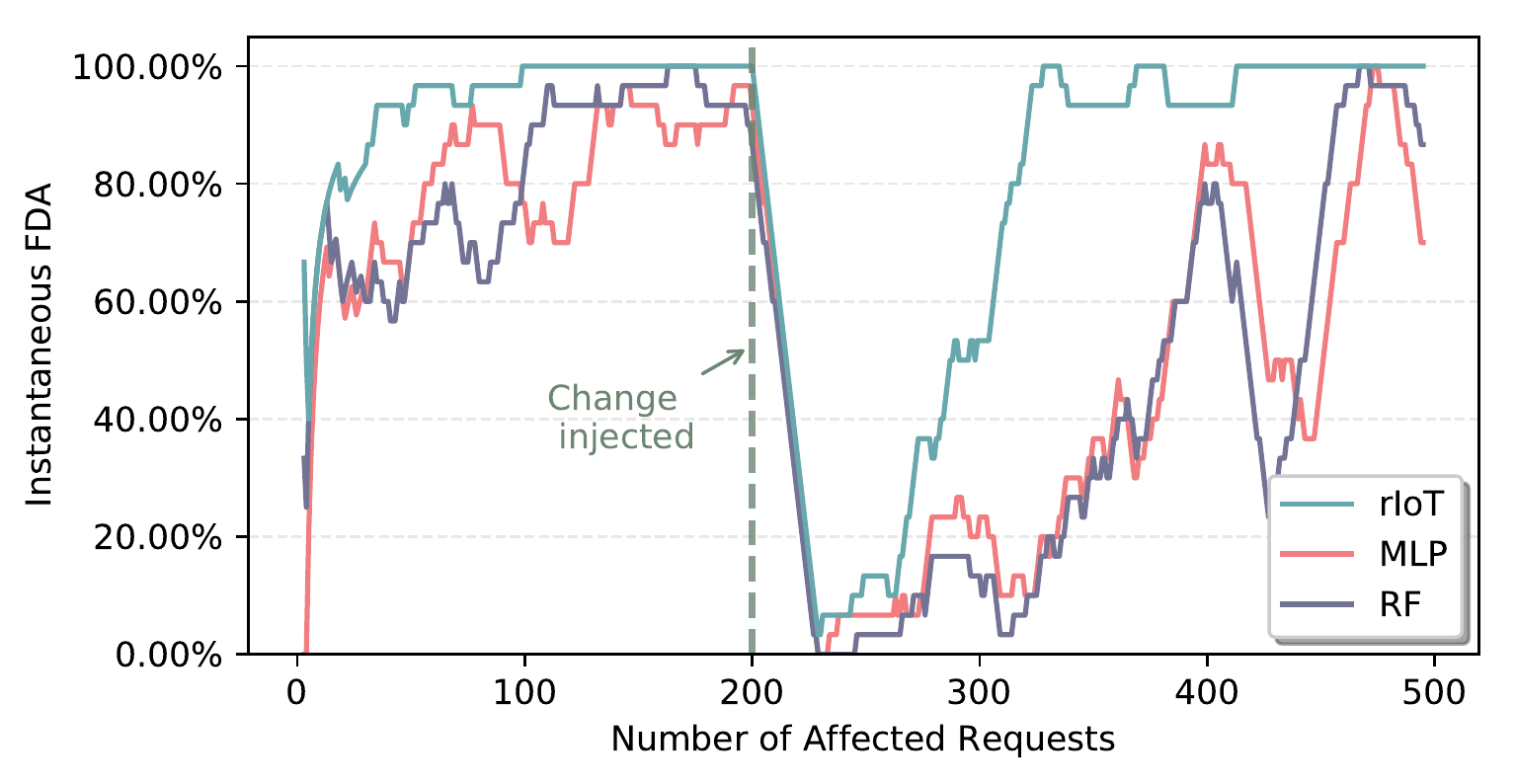}
\caption{{\it Experiment 4.} This figure demonstrates that \systemname can rebound from dramatic changes. The graph shows the FDA of \systemname, MLP, and RF when the user completely swaps two IoT devices in her environment after 200 requests. \systemname is much more agile in responding to the change.}
\label{fig:sim_envChange}
\end{figure}

To test responsiveness to dynamics, we simulated the situation when the user moves the devices in her home by switching the dining room light and the doorway light midway through the experiment, after all algorithms have initially converged. We use the average of the \textbf{FDA} of the previous 30 \request{}s to represent the instantaneous accuracy at this request; recall that the {\bf FDA} is the probability that this request is satisfied by each algorithm's first decision. Fig.~\ref{fig:sim_envChange} shows the results. 

The accuracy of all three approaches drops dramatically immediately following the switch. \systemname recovers significantly faster and restabilizes to a high accuracy. This quick recovery is because \systemname recomputes the \utility value of the affected models if it detects a disparity of between recent feedback and the previous learned model. Although it may seem faster, clearing the model and learning from scratch would harm the accuracy of unaffected requests. Further, there is no oracle that indicates that a dramatic change has occurred and thus a reset is in order. To the best of our knowledge, existing context-aware approaches do not consider environmental change explicitly, and thus must either retrain the entire model or wait until the new interactions dominate the old ones.

\subsection{\systemname's Feasibility on Mobile Devices}
We implemented \systemname on Android to demonstrate its feasibility on real IoT devices. All of our measurements are made using a Moto X (2nd Gen.) with Android 5.1 Lollipop.
% \subsection{Android implementation}
% \note{Optional: android implemenation with the testbed setup. No results.}

In automating device interactions, \systemname does incur overhead. We measured both the response latency and the feedback latency in \systemname. The former is the time between a user issuing a request and \systemname responding with a proposal. The feedback latency is measured as the time between when \systemname receives a piece of feedback until it finishes updating the local utility models. We used the same requests and context snapshots as in Section~\ref{sec:exp_realworld}. To test how the latency is related to the number of contexts, we added random numerical contexts to the original contexts; the number of total contexts is increased from 3 to 33; measurements are averaged over all requests.

\begin{figure}[t]
\centering
\includegraphics[width=.85\columnwidth]{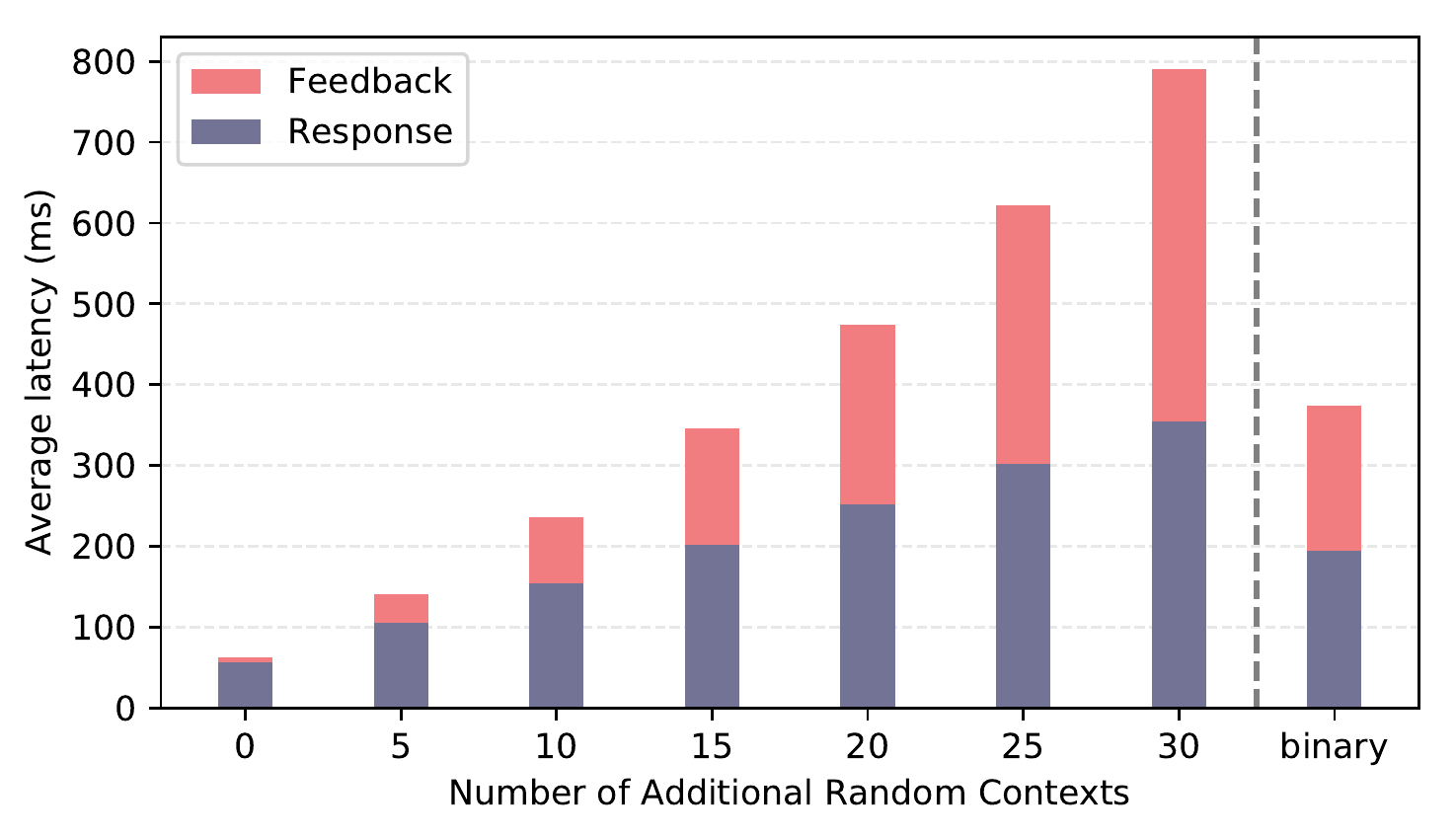}
\caption{{\it Experiment 5.} This figure shows the average response and feedback latency per request given an increasing number of contexts. The ``binary'' bar use 36 context attributes with only two possible values. In all cases, but especially for realistic numbers of contexts, \systemname is feasible on real devices.}
\label{fig:android}
\end{figure}

Fig.~\ref{fig:android} shows the results. Response latency blocks the UI; it is below 400ms even with the maximal number of contexts. Incorporating feedback into the models can be scheduled in the background, but even so, the overhead is reasonable. Note that increasing the number of contexts does not always increase the latency; for non-random contexts (e.g., the binary types in Fig.~\ref{fig:android}), \systemname learns the model more quickly, so requests and feedback can be processed more efficiently.
%The last bar(``binary'') shows the result for 36 binary contexts\footnote{The 36 binary contexts are two groups of binary state of the 18 motion sensors corresponding to the two categorical context. } and the \textit{Time} context. The reason why the latency for these 37 contexts is less is that 30 of the previous 33 contexts are randomly generated with no information. And \systemname will keep trying to learn knowledge from these random context. If all the \context{}s are informative, \systemname finds the decision boundry in the early rounds and thus has less overall computation.

% \begin{table}[t]
% \centering
% \small
% \caption{Average latency}
% \vspace{-.25cm}
% \label{tab:latency}
% \begin{tabularx}{\columnwidth}{|X|X|X|}
% \hline
% {\bf Number of contexts} & {\bf Response Latency (ms)} & {\bf Feedback Latency (ms)} \\\hline \hline
%  {3}  & 28.74  & 2.27           \\ \hline
% {37}  & 195.14 & 178.94         \\ \hline
% \end{tabularx}
% \end{table}
\section{Conclusion and Future Work}\label{sec:conclusion}
In this paper, we demonstrated \systemname's ability to provide context-aware automation and personalized human-device interaction with no setup cost from the user. \systemname ensures user-oriented decision making by explicitly including and responding to a user's immediate feedback. We evaluated \systemname on both real-world noisy data and simulated scenarios to show that \systemname performs well in terms of its accuracy and is resilient to environmental changes, which are common in the mobile IoT world. \systemname is also sufficiently lightweight and efficient to run on mobile devices. 
However, work with \systemname is not complete.
%Security and privacy are essential for the IoT. In the current form of \systemname, we rely on external source to provide authentication and authorization when establish connections to the IoT devices. In the future, \systemname can potentially provide context-aware anomaly detection or authentication based on the user's behavior pattern. 
For instance, our presentation of \systemname assumes that each request corresponds to one action on one device. In practice, a user may expect multiple devices to act together. An extension of \systemname could allow devices to coordinate to submit shared proposals. In addition, interfacing \systemname with existing smart home platforms like Google Smart Home\footnote{https://developers.google.com/actions/smarthome/} is also worth researching.
%A simple solution to this is that \systemname can return several proposals with \utility higher than a threshold. In the future, more intelligent selection of device combination can be studied. 
In conclusion, \systemname opens new possibilities of an IoT world that is truly personalized, providing users seamless and intuitive interactions with the digitized world around them.

% We also assume that the user does not make mistakes when giving request and feedback which may not always be true in real life.  
% \note{security and privacy}
% \note{Smart exploration of the utility function.}
% \note{include user error}

% We quantitatively evaluated \systemname on both real-world datasets and simulated scenarios with inaccurate sensor readings and environmental change. The result shows that \systemname has better performance in terms of accuracy and learning speed compared to the two machine learning algorithms and can recover faster from the environmental change. 

\section*{Acknowledgments}
This material is based upon work supported by the National Science Foundation under Grant No. CNS-1703497. %Any opinions, findings, and conclusions or recommendations expressed in this material are those of the author(s) and do not necessarily reflect the views of the National Science Foundation.  
%Any opinions, findings, and conclusions or recommendations expressed in this material are those of the author(s) and do not necessarily reflect the views of the National Science Foundation.

%\begin{acks}
%{\color{red}acknowledgements... not in double blind, tho...}

%The work is supported by the %\grantsponsor{GS501100001809}{National
%  Natural Science Foundation of
%  China}{http://dx.doi.org/10.13039/501100001809} under %Grant
%No.:~\grantnum{GS501100001809}{61273304\_a}
%and~\grantnum[http://www.nnsf.cn/youngscientists]{GS5011000%01809}{Young
%  Scientists' Support Program}.

%\end{acks}

% Bibliography
\bibliographystyle{plain}
\bibliography{main}

\end{document}